\documentclass[12pt]{article}
\usepackage{ifpdf}
\ifpdf
\usepackage{graphicx,color}
\usepackage{hyperref}
\else
\usepackage[dvipdfmx]{graphicx,color}
\usepackage[dvipdfmx]{hyperref}
\fi
\usepackage{amssymb,amsfonts,amsmath,cancel,cite,multirow}
\usepackage[capitalise]{cleveref}

\newcommand{\ave}[1]{\left \langle #1 \right \rangle}

\newcommand{\pdiff}[3]{
\if 1#1   \frac{\partial #2 }{\partial #3 }
\else  \frac{\partial^#1#2 }{\partial #3^#1 } \fi}

\newcommand{\diff}[3]{
\if 1#1  \frac{{\rm d} #2 }{{\rm d} #3 }
\else  \frac{{\rm d}^{#1} #2 }{{\rm d}#3^{#1} } \fi
}

\newcommand{\Lam}{\Lambda}

\newcommand{\Del}{\Delta}
\newcommand{\ovl}{\overline}

\newcommand{\si}{\sigma}

\newcommand{\curlS}[0]{\mathcal{S} }

\newcommand{\eqs}[1]{\begin{equation}\begin{split} #1 \end{split}\end{equation}}

\begin{document}
\begin{titlepage}
\hfill IPMU~16-0165 \\
\begin{flushright}
\end{flushright}

\vskip 1.35cm
\begin{center}

{\large
{\bf
Light Stops, Heavy Higgs, and Heavy Gluinos \\
in Supersymmetric Standard Models with Extra Matters
}}
\vskip 1.2cm

Junji Hisano$^{a,b,c}$,
Wataru Kuramoto$^b$,
and
Takumi Kuwahara$^b$\\

\vskip 0.4cm

{\it $^a$Kobayashi-Maskawa Institute for the Origin of Particles and the Universe (KMI),
Nagoya University, Nagoya 464-8602, Japan}\\
{\it $^b$Department of Physics,
Nagoya University, Nagoya 464-8602, Japan}\\
{\it $^c$
Kavli IPMU (WPI), UTIAS, University of Tokyo, Kashiwa, 277-8584, Japan}
\date{\today}

\vskip 1.5cm

\begin{abstract}
We have explored the possibilities of scenarios with heavy gluinos and light stops in supersymmetric (SUSY) standard models with extra vector-like multiplets.
If we assume the hierarchical structure for soft masses of the minimal supersymmetric standard model (MSSM) scalar fields and extra scalars, the light stop and the observed Higgs boson can be realized.
While the stau is the lightest SUSY particle (LSP) in broad parameter space,
we have found that the neutralino LSP is realized in the case that the non-zero soft parameters for the MSSM Higgs doublets or the non-universal gaugino masses are assumed.
\end{abstract}

\end{center}
\end{titlepage}

\section{Introduction \label{sec:intro}}

Supersymmetry (SUSY) is one of the promising extensions of the standard
model (SM).  New particles, with opposite spin statistics to the
SM particles, are naturally introduced by extending spacetime to 
one with Grassmann coordinates.  If these new particles, called
sparticles, lie around the TeV scale, SUSY provides us with some
phenomenological implications.  The lightest supersymmetric particle
(LSP) may explain the dark matter abundance in the universe.  The gauge coupling
unification also works well, and is compatible with grand
unification theories (GUTs).

The ATLAS and CMS collaborations found a scalar boson consistent with
the SM Higgs boson \cite{Aad:2012tfa,Chatrchyan:2012ufa}, and reported
that its mass is around 125~GeV \cite{Aad:2015zhl}. In the minimal
supersymmetric standard model (MSSM), the light Higgs boson mass is bounded
from above at tree level. To explain the observed Higgs boson mass,
several ideas have been proposed: the introduction of large quantum
corrections to the light Higgs mass, adding vector-like extra
matters \cite{Babu:2008ge,Martin:2009bg}, pushing up the SUSY
breaking scale \cite{Giudice:2011cg,Ibe:2011aa}, realizing large
A-term or next-to-MSSM \cite{Hall:2011aa}, and so on.

There is also no signal of sparticles and no significant
deviation from the SM predictions at the LHC experiments (\textit{e.g.},  see Refs.~\cite{Aad:2014wea,Chatrchyan:2014lfa}).  In particular, the
masses of new colored particles are severely constrained; for
instance, gluinos should be heavier than about 1.9~TeV in a simplified
mass spectrum \cite{Adam:2016ICHEP}.

Heavy gluinos na\"ively indicate heavy squarks at the low-energy
scale.  In fact, in order to obtain the heavy gluinos at the low-energy
scale, we require a large value for the gluino mass at an initial
scale (such as the GUT scale $\sim 10^{16}~\text{GeV}$ or the Planck
scale $\sim 10^{18}~\text{GeV}$). According to renormalization
group equation (RGE) analysis, heavy gluinos at the input scale
lead to heavy squarks and a large A-term at the one-loop order
in the MSSM.  The heavy stop is unfavorable from the
naturalness point of view since it requires fine-tuning between the
soft mass and the supersymmetric mass for the up-type MSSM Higgs
doublet.

There are other things to be considered in the supersymmetric extended
models: SUSY flavor problems.  To suppress the
flavor-changing neutral current (FCNC) processes, it is required that
the sfermion masses for the first two generations are degenerate,
decoupled, and/or aligned. Many models have been constructed assuming
that SUSY breaking is mediated by gauge interactions, since it gives a
flavor-blind structure for sfermion masses.  However, in those
flavor-blind SUSY-breaking scenarios in the MSSM, it is difficult to
discover sparticles at the LHC.  In fact, as we mentioned, the
Higgs mass constraint leads to large squark and gluino masses in
such scenarios.

In this paper, we consider the introduction of vector-like extra matters to the MSSM.
Considering the two-loop RGE effects from the gauge interactions, the soft masses for the extra scalar fields give a negative contribution to those for MSSM sfermions.
Thus, if the soft masses for the extra scalar fields are somewhat larger than those for the MSSM sfermions at the initial scale, the physical masses for the MSSM sfermions become smaller.
Since the low-energy A-terms are not affected in the presence of the extra matters, the A-terms are effectively larger than the soft masses for the MSSM sfermions.
As a result, the light Higgs mass gets a large radiative correction, so that the observed Higgs mass is realized.
The FCNC processes may be suppressed if the choosen sfermion masses are zero at the initial scale.
Thus, our setup is consistent with various observations while lighter sparticles may be predicted.

A hierarchical structure for soft masses between the MSSM and
vector-like extra matters may be realized in the context of gaugino
mediation scenarios (\textit{e.g.},
Refs.~\cite{Chacko:1999mi,Schmaltz:2000gy,Cheng:2001an}). In these
scenarios, only gauginos couple to the SUSY-breaking brane, and the
sfermions in the MSSM feel the SUSY breaking via the gaugino loops. If
the vector-like extra matters also couple with the SUSY-breaking
brane, their soft masses may be larger than the gauginos at the input
scale since the gaugino masses are suppressed by the $U(1)_R$
breaking.  As a result, smaller sfermion masses are expected, as
explained above, even if the gauginos have a mass of several TeV due
to the additional negative contribution.  Here, the MSSM Higgs
multiplets may be coupled with the SUSY-breaking brane or they may
not. The soft terms for the MSSM Higgs doublets are model dependent.

At first glance, this setup looks to include fine-tuning since the two-
and one-loop contributions to sfermion masses are comparable with each
other. This might come from some dynamics in which the gaugino masses
are suppressed by one-loop factors.
In addition, if the two-loop contribution is much larger than the one-loop one,
such theories are not for our universe since the vacua would break color and/or charge.
If the probability distribution for the soft masses of the extra matters is an increasing function,
our proximity to the tachyonic boundary may be understood in the context of the anthropic principle.

Similar work has been done in the context of composite
supersymmetric models \cite{Cleary:2015koc}.  We suggest another
picture for the scenarios with heavy gluino and light stop, in which
the perturbative description works well until the GUT or the Planck
scale.

The organization of this paper is the following: in \cref{sec:model},
we will discuss our models—their particle contents and initial conditions for
soft parameters.  Next, we will give the numerical results on the
light Higgs mass and the light stop mass in our models.  In this study
we use the RGEs at two-loop level, as shown in \cref{app:RGE}.
Finally, we conclude our study in \cref{sec:conclusion}.

\section{Model \label{sec:model}}
First of all, we briefly give the details of our model.  Throughout this
paper, we consider supersymmetric models with vector-like extra matters, as
mentioned in \cref{sec:intro}.  If they are in $SU(5)$ irreducible
representations (\textit{e.g.}, $\mathbf{5}+\ovl{\mathbf{5}}$ or
$\mathbf{10}+\ovl{\mathbf{10}}$ ), the unification of gauge couplings
is maintained \cite{Martin:2009bg}.  For simplicity, we assume that
the extra matters do not mix with the MSSM fields.

The superpotential for our model is given by
\eqs{ W & = (Y_u)_{ij}
  \ovl U_i Q_j H_u - (Y_d)_{ij} \ovl D_i Q_j H_d - (Y_e)_{ij} \ovl E_i
  L_j H_d + \mu_H H_u H_d + \Del W_{\text{add}}\,.
\label{superpotential}}
where the first four terms are the ordinary MSSM superpotential. Here, we
suppress the gauge indices.  $Q_i$ and $L_i$ are the $SU(2)_L$ doublet
chiral superfields including left-handed quarks and leptons, while
$\ovl U_i$, $\ovl D_i$, and $\ovl E_i$ are the $SU(2)_L$ singlet
chiral superfields including the up-type quark, down-type quark, and
charged lepton, respectively.  The MSSM Higgs chiral superfields are
denoted by $H_u$ and $H_d$.  The subscripts $i,j=1,2,3$ denote the
generations, and $Y_u$, $Y_d$, and $Y_e$ are the $3 \times 3$ Yukawa
matrices.

The last term, $\Del W_{\text{add}}$, in Eq.~(\ref{superpotential}) is
the additional superpotential with vector-like extra matters. In the
$\mathbf{5}+\ovl{\mathbf{5}}$ extension, we introduce a pair of $
\mathbf{5} = (D', \ovl L') $ and $ \ovl{\mathbf{5}} = (\ovl D', L')$,
and the superpotential $\Del W_{\text{add}}$ contains supersymmetric
mass terms for the extra matters. The further extension is
straightforward.

The soft SUSY-breaking terms in our setup are
\eqs{
- \mathcal{L}_{\text{soft}} & = \frac12 M_3 \widetilde g \widetilde g + \frac12 M_2 \widetilde W \widetilde W + \frac12 M_1 \widetilde B \widetilde B + \text{c.c.} \\
& + (A_u)_{ij} \widetilde{\ovl u}_i \widetilde q_j H_u
- (A_d)_{ij} \widetilde{\ovl d}_i \widetilde q_j H_d
- (A_e)_{ij} \widetilde{\ovl e}_i \widetilde l_jH_d + \text{c.c.} \\
& + (m_Q^2)_{ij} \widetilde q^\dag_i \widetilde q_j
+ (m_L^2)_{ij} \widetilde l^\dag_i \widetilde l_j
+ (m_{\ovl U}^2)_{ij} \widetilde{\ovl u}_i \widetilde{\ovl u}^\dag_j
+ (m_{\ovl D}^2)_{ij} \widetilde{\ovl d}_i \widetilde{\ovl d}^\dag_j
+ (m_{\ovl E}^2)_{ij} \widetilde{\ovl e}_i \widetilde{\ovl e}^\dag_j\\
& + m_{H_u}^2 H_u^\dag H_u + m_{H_d}^2 H_d^\dag H_d + (b H_u H_d + \text{c.c.})
- \Del \mathcal{L}_{\text{soft:add}}\, .
}
The objects with a small letter and a tilde ($\widetilde q_i,
\widetilde{\ovl u}_i, \widetilde{\ovl d}_i, \widetilde l_i$, and
$\widetilde{\ovl e}_i$) correspond to the superpartners of the SM
fermions, while $\widetilde g$, $\widetilde W$, and $\widetilde B$ are
respectively gluinos, winos, and binos, which are the fermionic partners of
gluons, weak bosons, and the hypercharge gauge boson.  We use the same
letters, $H_u$ and $H_d$, for scalar components of the MSSM Higgs
doublets.  The terms (except for the last one) correspond to soft terms in
the MSSM; the gaugino masses $M_{1,2,3}$, the scalar trilinear
coupling matrices $A_{u,d,e}$, the holomorphic Higgs soft mass $b$,
the non-holomorphic soft masses for Higgs doublets $m_{H_u}^2$ and
$m_{H_d}^2$, and the non-holomorphic soft masses for sfermions $ m_i^2 ~ (i =
Q, L, \ovl U, \ovl D, \text{and} ~\ovl E)$.  $\Del
\mathcal{L}_{\text{soft:add}}$ denotes the soft SUSY-breaking term for
the extra matters.  For the $\mathbf{5}+\ovl{\mathbf{5}}$ extension,
the additional term is given by
\eqs{ - \Del
  \mathcal{L}_{\text{soft:add}} & = m_{L'}^2 \widetilde l'^\dag
  \widetilde l' + m_{\ovl L'}^2 \widetilde{\ovl l'} ~ \widetilde{\ovl
    l'}^\dag + m_{D'}^2 \widetilde{d'}^\dag \widetilde{d'} + m_{\ovl
    D'}^2 \widetilde{\ovl d'} \widetilde{\ovl d'}^\dag \, .  }
In this work we assume that the soft parameters are given at the GUT
scale ($=2\times10^{16}~\text{GeV}$). At this scale, the
non-holomorphic sfermion masses and the A parameters in the MSSM 
vanish, while the gaugino masses and also the soft terms for the
extra matters are non-zero. The non-holomorphic and holomorphic soft
masses for the MSSM Higgs doublets are model dependent. When we impose
the GUT relations for the gaugino masses and the soft terms for the
extra matters, they are given as
\eqs{
M_3 & = M_2 = M_1 = M_{1/2}\, , \\
m_{\ovl L'}^2 & = m_{D'}^2 \, , ~~~
m_{L'}^2 = m_{\ovl D'}^2 \, .
}
In the following phenomenological analysis, we show the results
with and without the GUT relations.

In \cref{app:RGE}, we show the modification of the RGEs in the
presence of the extra matters.  Here, we give the approximate analytic
solutions of RGEs for soft masses in the first and second generations.
The RGE for the soft masses $m_s^2$ is
\eqs{
  \diff{1}{m_s^2}{\ln\mu} = \sum_{A=1,2,3} \left[- \frac{1}{16\pi^2} 8 g_A^2 C_A(s) |M_A|^2
  + \frac{1}{(16\pi^2)^2} 4 g_A^4 C_A(s) \sum_r 2 S_A(r) m_r^2 \right]\, .
\label{eq:RGEsforsfermion}
}
Here, $C_A(s)$ and $S_A(s)$ denote the quadratic Casimir invariant and
the Dynkin index for the chiral multiplet $s$. The summation in the
bracket is dominated by the contribution from the extra matters in our
setup. The Yukawa coupling constants are negligible in this equation.
Assuming $m_s^2=0$ at the initial scale $\Lam$, the approximate
solution is given by
\eqs{
m_s^2(\mu) = \sum_{A=1,2,3}\frac{2 C_A(s)}{b_A} \left[ |M_A(\Lam)|^2 - |M_A(\mu)|^2 - \frac{\alpha_A(\Lam) - \alpha_A(\mu)}{4\pi} \sum_r 2 S_A(r) m_r^2\right] \, ,
}
where $b_A$ is the one-loop coefficient of the $\beta$-function for gauge coupling $g_A$.
It turns out that the condition for no tachyonic sfermion is
\eqs{
|M_A(\Lam)|^2 \left( 1 + \frac{\alpha_A(\mu)}{\alpha_A(\Lam)} \right)
> \frac{\alpha_A(\Lam)}{4\pi} \left( \sum_r 2 S_A(r) m_r^2 \right) \, .
}
Here, we use the fact that $(\alpha_A(\Lam) - \alpha_A(\mu))/b_A > 0$
regardless of whether the corresponding gauge interaction is
asymptotically free or not.  This condition implies that the large
mass hierarchy between the gauginos and vector-like extra matters
leads to the light sfermions even if the gauginos are much larger than
1~TeV.  To be more concrete, if the soft masses of the extra matters are
about ten times larger than those of gauginos, the dominant
contribution to the sfermion mass is approximately cancelled.

The squared masses for the third-generation sfermions are reduced more than
those for the first two generation sfermions due to the one-loop
contribution from the Yukawa couplings to the RGEs. This is plausible
from the naturalness point of view. The stop masses in the RGE for $m_{H_u}^2$ are smaller, so that the absolute value of $m^2_{H_u}$ is
smaller. This means that the fine-tuning between the
supersymmetric mass and soft masses in the Higgs potential is relaxed.

In the next section we show some phenomenological studies for our setup. There
we only introduce a pair of $\mathbf{5}+\ovl{\mathbf{5}}$
multiplets in our numerical analysis.  If many pairs of
$\mathbf{5}+\ovl{\mathbf{5}}$ and 
$\mathbf{10}+\ovl{\mathbf{10}}$ are introduced and the constrained MSSM
spectrum is assumed, the observed Higgs mass is realized as the
framework of the large A-term scenario even without large soft masses for
extra matters \cite{Moroi:2012kg,Moroi:2016ztz}, though the stop masses are
heavier.  Introduction of the larger multiplets or many fields would make
our points unclear, so we consider the case of a pair of
$\mathbf{5}+\ovl{\mathbf{5}}$ multiplets.

\section{Numerical Results\label{sec:num}}

In this section, we present our numerical results for the light Higgs mass
and the light stop mass in the SUSY SM with a pair of
$\mathbf{5}+\ovl{\mathbf{5}}$ multiplets.

Before we show our numerical results, we briefly summarize our
procedure to evaluate the low-energy mass spectrum.  We give initial
conditions for soft parameters at the GUT scale ($ = 2\times
10^{16}~\text{GeV}$), and we evolve the soft parameters with the RGEs
at the two-loop level \cite{Martin:1993zk}.  We also obtain the gauge couplings
and Yukawa couplings at the GUT scale by using the two-loop RGEs for
them \cite{Machacek:1983tz,Machacek:1983fi}.  We set the SUSY-breaking
scale to be 1~TeV, and then we treat the effective theories above the
SUSY-breaking scale as the SUSY SM with extra matters.  The modification
of RGEs due to extra matters is shown in \cref{app:RGE}.

Since the soft parameters for the MSSM Higgs doublets $m_{H_u}^2$ and
$m_{H_d}^2$ at the SUSY-breaking scale are determined by the RGE
evolution, we evaluate the supersymmetric higgsino mass $\mu_H$ and
the holomorphic Higgs soft mass $b$-terms via the conditions for
potential minima,
\eqs{
|\mu_H|^2 & =
\frac{1}{1-\tan^2\beta} \left[ \tan^2\beta \left( m_{H_u}^2 + \frac{1}{2 v_u} \pdiff{1}{\Del V}{v_u} \right) - \left( m_{H_d}^2 + \frac{1}{2 v_d} \pdiff{1}{\Del V}{v_d} \right) \right] - \frac{m_Z^2}{2} \, , \\
\frac{2 b}{\tan2\beta} & = \left( m_{H_u}^2 + \frac{1}{2 v_u}
  \pdiff{1}{\Del V}{v_u} \right) - \left( m_{H_d}^2 + \frac{1}{2 v_d}
  \pdiff{1}{\Del V}{v_d} \right) - m_Z^2 \cos 2\beta \, .  }
Here, $v_{u,d} = \ave{H_{u,d}^0}$ are the vacuum expectation values
(VEVs) for the neutral components of the MSSM Higgs doublets, and
$\tan\beta = v_u/ v_d$ is the ratio of VEVs.  $\Del V$ denotes the
one-loop effective potential.  Then, we obtain the low-energy Higgs
mass with the use of \texttt{SPheno} \cite{Porod:2003um}.  In this
study, we do not include finite threshold corrections to the scalar
soft parameters \cite{Hisano:2000wy}, though we use the two-loop RGEs
for them.  This is because the finite corrections are expected to
be a subdominant contribution and strongly depend on the supersymmetric
mass parameters for the vector-like extra matters.

First, we assume the GUT relations for soft parameters as input
parameters.  To be more concrete, we impose the no-scale type initial
conditions for only the MSSM multiplets, that is, the soft parameters are
zero except for the gaugino masses and $b$-term at the GUT scale.
Furthermore, we also impose the initial conditions for the gaugino
masses and soft scalar masses of extra matters to be degenerate, and
their masses are denoted by $M_{1/2}$ and $ m_{\text{vec}}^2$,
respectively. Unless we mention otherwise, the soft masses for the MSSM Higgs
doublets are set to be zero. We take $\tan\beta = 10$
and $\mu_H>0$ for simplicity in this paper.

\begin{figure}[t]
\centering
\includegraphics[width=7cm,clip]{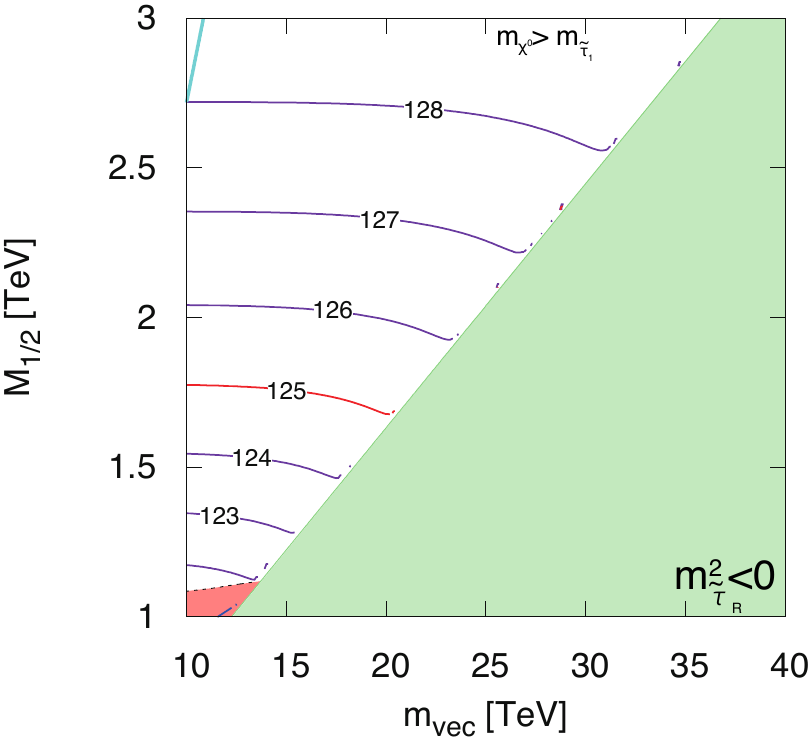}
\includegraphics[width=7cm,clip]{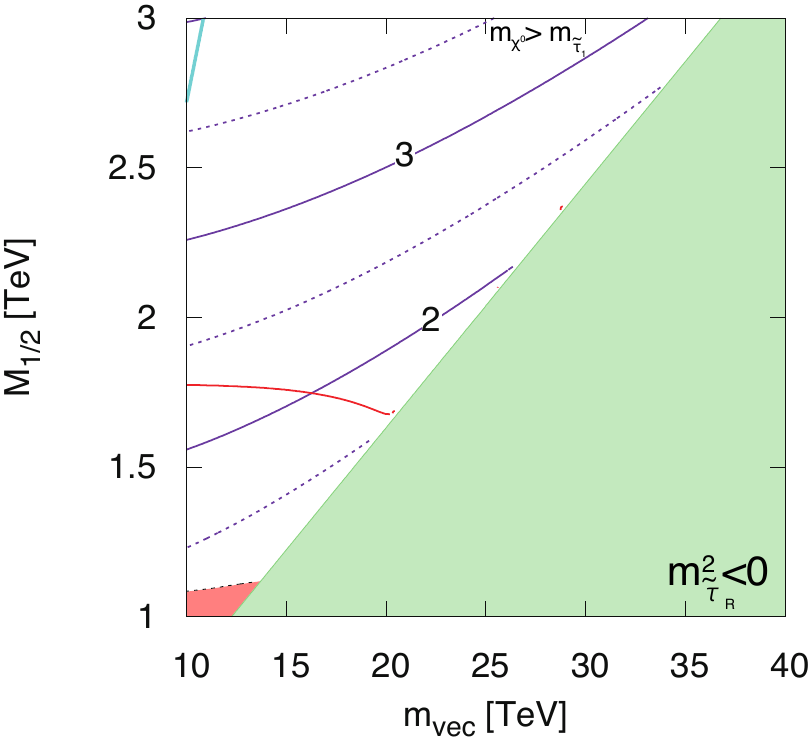}
\caption{ Dependences of light Higgs mass (left) and light stop mass
  (right).  An additional pair of $\mathbf{5}+\ovl{\mathbf{5}}$ is
  introduced as extra matter.  The red solid line in each figure
  corresponds to the 125 GeV Higgs mass in our scenario.  The green shaded
  region is excluded by tachyonic stau constraint.  The gluino mass is
  below 1.9~TeV in the red shaded region.  The bino-like neutralino is the LSP
  above the cyan solid line, while the light stau is the LSP below the
  line. }
\label{fig:55bar}
\end{figure}

In \cref{fig:55bar}, we show the numerical results of the light
Higgs mass (left panel) and the light stop mass (right panel) in
the aforementioned setup.  The numbers in these panels indicate the Higgs
mass in GeV (left panel) and the light stop mass in TeV (right
panel).  The observed Higgs mass (125 GeV) is realized on the red solid
line in each panel.  The green shaded region corresponds to the case
that the stau becomes tachyonic, and thus the region is excluded due
to the charge-breaking minima.  The red shaded region in the
bottom left side of each figure shows that the mass of gluinos is below
1.9~TeV.  The cyan solid line illustrates the boundary where the LSP
is changed; the LSP is the bino-like neutralino above the line while
it is the light stau below the line.

In this setup, the soft masses for all scalars in the MSSM receive a negative
contribution from heavy extra matters.
The positive contributions from gaugino masses to sleptons are smaller
than those to squarks since $ M_{1,2} $ and $ g_{1,2} $ are smaller
at the low-energy scale.  As a result, the right-handed stau may be
the lightest sparticle and become tachyonic.  Here, we show only the
most stringent constraint (tachyonic stau in this figure).
As we will see below, there are other constraints such as the
color-charge breaking vacuum (tachyonic stop: $m_{\widetilde t}^2 <
0$) or no electroweak symmetry breaking (EWSB: $m_{H_u}^2 > 0$).
Such constraints in \cref{fig:55bar} hide behind the constraint from the
tachyonic stau.  \cref{fig:55bar} shows that we cannot explain the
125~GeV Higgs boson with a stop below 1.5~TeV in this setup.  This is
because the stau is tachyonic when the 1~TeV stop is realized.

\begin{figure}[t]
\centering
\includegraphics[width=7cm,clip]{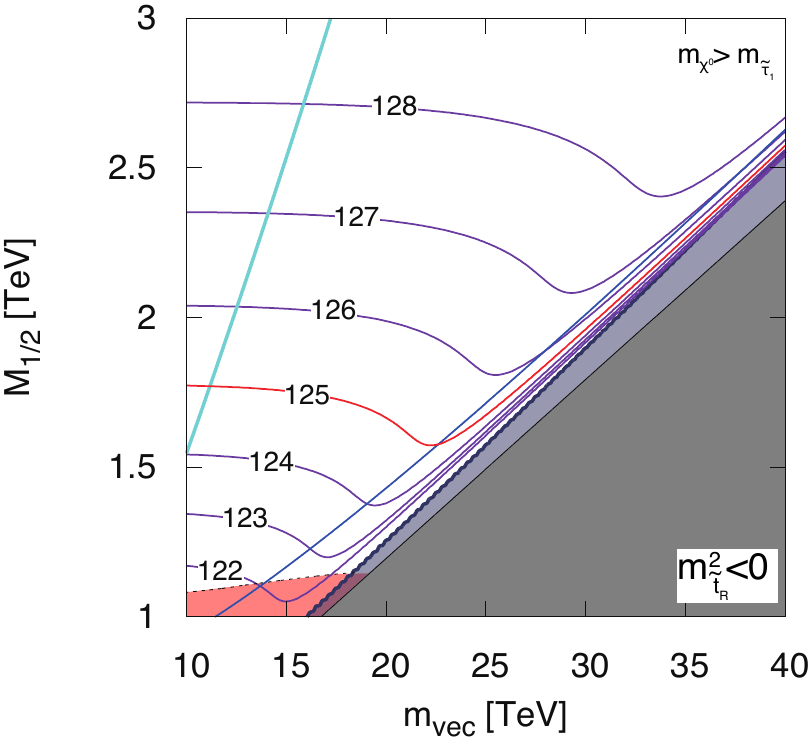}
\includegraphics[width=7cm,clip]{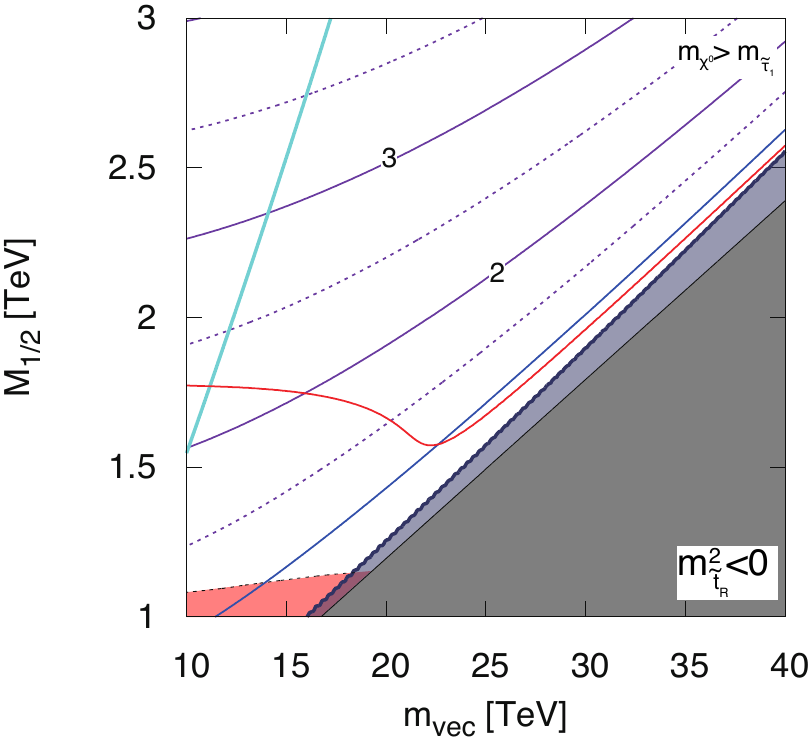}
\caption{ Dependences of light Higgs mass (left) and light
  stop mass (right).  An additional pair in
  $\mathbf{5}+\ovl{\mathbf{5}}$ is introduced.  Colored scalars in
  extra multiplets have large soft mass while non-colored ones do not.
  The red solid line corresponds to the 125 GeV Higgs boson in our
  scenario. The gray shaded region is excluded by the constraint for
  tachyonic stops. 
  The stop is the LSP in the navy shaded region. 
  the gluino mass is below 1.9~TeV in the red shaded region.}
\label{fig:55bar_hier}
\end{figure}

Next, in order to avoid the tachyonic stau, we assume that the colored
scalars in the extra multiplets obtain non-zero soft scalar masses, while the
soft masses for the non-colored ones are zero or negligible
at the GUT scale, as
\eqs{
  m_{L'}^2 = m_{\ovl L'}^2 = 0 \, , ~~~~~~ m_{D'}^2 = m_{\ovl D'}^2 =
  m_{\text{vec}}^2 \, .
\label{eq:hier_input}
}
The sleptons do not get much negative contribution from the extra
matters by imposing this condition. In \cref{fig:55bar_hier}, we show
the numerical results of the light Higgs and the light stop masses
under this condition.

In this case, the tachyonic stop gives a severe constraint.  The gray 
shaded area is excluded due to the tachyonic stop. 
The stop is the LSP in the navy shaded region. 
The blue solid lines in both figures correspond to the light stop with a mass of
1~TeV.  Near the boundary of the charge-color breaking vacuum, the
large A-term is realized due to the large input value for the gaugino
mass $M_{1/2}$, and then the radiative correction to the Higgs boson
mass is enhanced.  Since the A-term, however, becomes somewhat larger
than the stop mass as the value $m_{\text{vec}}$ approaches the
boundary, the observed Higgs mass cannot be realized. Indeed, the radiative
correction to the light Higgs mass is maximum when $A_t/m_{\widetilde t} \sim \sqrt6$. 
The larger A-term suppresses the radiative
correction. Thus the light Higgs mass contour lines are parallel to
each other near the tachyonic stop boundary.

It is known that the large A-terms lead to new deep charge-color
breaking (CCB) minima \cite{Komatsu:1988mt}. 
Near the stop LSP boundary, the deep CCB minimum appears.
Above the boundary, the simplified condition 
(for example, see Refs. \cite{Komatsu:1988mt,Casas:1995pd})
\eqs{
|(A_u)_{33}|^2 \leq 3 ((m_Q^2)_{33} + (m_{\ovl{U}}^2)_{33} + m_{H_u}^2 + \mu_H^2)
}
is satisfied.  Here, all soft parameters are estimated at the
renormalization scale $\mu = 1~\text{TeV}$.  While there are other CCB
directions of the scalar potential, it is found that the CCB
directions for sbottom and stau do not particularly constrain the
parameter region. On the red solid line where the 125~GeV Higgs boson is realized, 
the deep CCB minimum does not appear. Even in the remaining 
results of this paper, the deep CCB minimum is irrelevant to explain the lightest Higgs mass.

\cref{fig:55bar_hier} shows that the light stop with a mass of less than 1~TeV and the observed mass for the light Higgs are compatible with this scenario.
Even though we reduced the negative contribution to the soft masses for the sleptons,
the LSP is still stau in this region.  While the stau (next-to-)LSP scenarios
are severely constrained if the $R$-parity is conserved, it is
possible to evade such constraints when the $R$-parity is
slightly broken.\footnote{For instance, see
  Refs.~\cite{Ito:2011xs,Evans:2016zau,Ishiwata:2008tp} for the collider phenomenology
  of the stau (N)LSP scenario, and Ref.~\cite{Ibe:2016gir} for the
  thermal leptogenesis scenario with $R$-parity violation and the
  gravitino LSP.}
Even if the $R$-parity is conserved, the stau NLSP
is still allowed in the axino DM scenario (see
Ref.~\cite{Choi:2011yf}).

Next, we consider the cases of avoiding the stau LSP scenario.
We devote the last part of this section to the following two cases: (1) positive non-zero input values for $m_{H_u}^2$ and $m_{H_d}^2$, and (2) non-universal gaugino masses.
In these cases, as we will see, we find the possibilities that the neutralino LSP and the light stop with mass about 1~TeV  are realized.

\begin{figure}[t]
\centering
\includegraphics[width=7cm,clip]{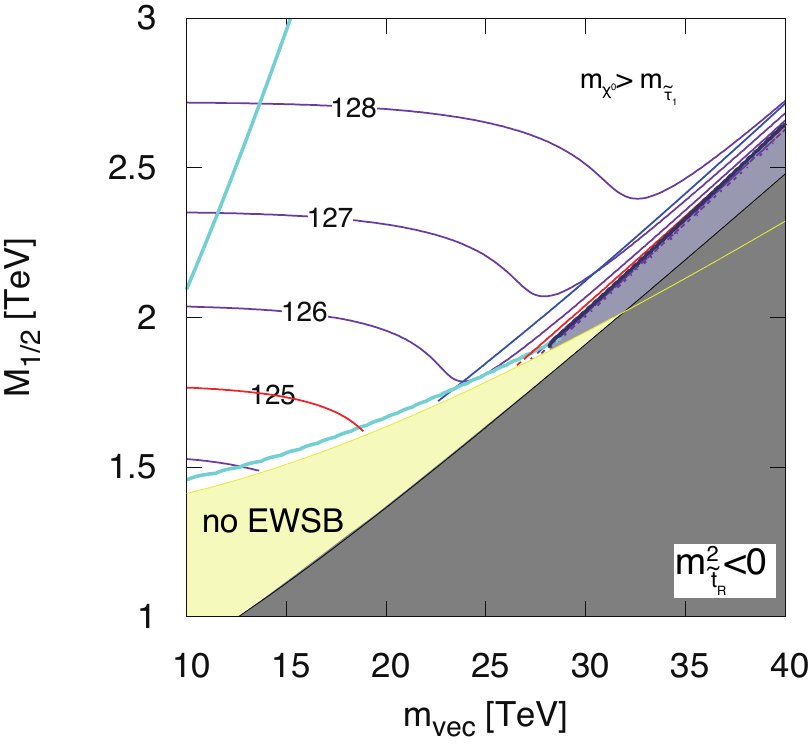}
\includegraphics[width=7cm,clip]{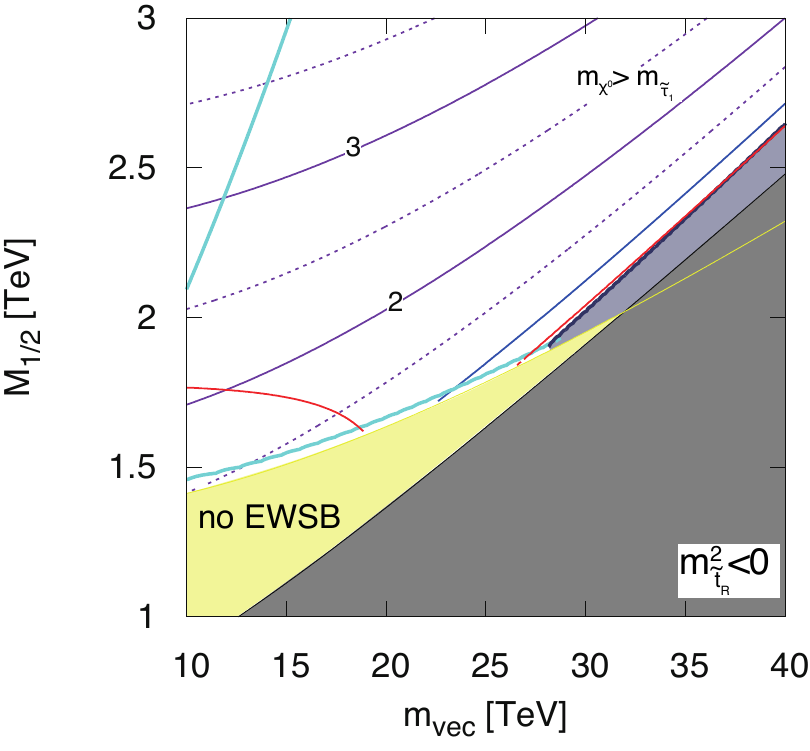}
\caption{ Dependences of light Higgs mass (left) and light
  stop mass (right).  An additional pair in
  $\mathbf{5}+\ovl{\mathbf{5}}$ is introduced.  Colored scalars in
  extra multiplets have large soft mass while non-colored ones do
  not. Positive initial values for the MSSM Higgs doublet masses are
  assumed ($m_{H_u}^2 = m_{H_d}^2 = (2.0~\text{TeV})^2$ at the GUT
  scale).  The red solid line corresponds to the 125 GeV Higgs mass in our
  scenario.  The gray shaded region is excluded by tachyonic stops, and
  the yellow shaded region is excluded due to no radiative EWSB. 
  The navy shaded region corresponds to the stop LSP.}
\label{fig:55bar_pos}
\end{figure}

Let us consider the case of positive $m_{H_u}^2$ and $m_{H_d}^2$
at the GUT scale.  As we mentioned in \cref{sec:intro}, the MSSM Higgs
doublets can have non-zero soft masses in the context of
gaugino mediation.  Imposing positive $m_{H_u}^2$ at the GUT scale
improves the fine-tuning between the soft and supersymmetric masses.
At the low-energy scale, we obtain small absolute values for
$m_{H_u}^2$, and then the higgsino-like neutralino is the LSP.

In \cref{fig:55bar_pos}, we show the numerical results for positive $m_{H_u}^2$ and $m_{H_d}^2$. Here, at the GUT scale,
\eqs{
m_{H_u}^2 = m_{H_d}^2 = (2.0~\text{TeV})^2 \, ,
}
while we set zeros for squarks and sleptons as in previous figures. We also assume the same condition for the extra matters as \cref{eq:hier_input}, so that
the tachyonic stop is the strongest constraint in the large $m_{\text{vec}}^2$ limit.
In addition to the tachyonic stop region, there is the region of no EWSB since the negative correction to $m_{H_u}^2$ via the RGEs is smaller; this is shaded yellow in \cref{fig:55bar_pos}. Two cyan lines illustrate the boundary where the LSP changes into the other sparticle.
The stau is the LSP in the medium region between the two cyan lines.
Above the upper line the bino-like neutralino LSP is realized, while the higgsino-like neutralino is the LSP below the bottom line. The stop is the LSP in the navy shaded region.
In the tiny region around the no EWSB boundary, the higgsino-like neutralino is the LSP, the observed Higgs mass, and the stop with a mass of about 1~TeV.

\begin{figure}[t]
\centering
\includegraphics[width=7cm,clip]{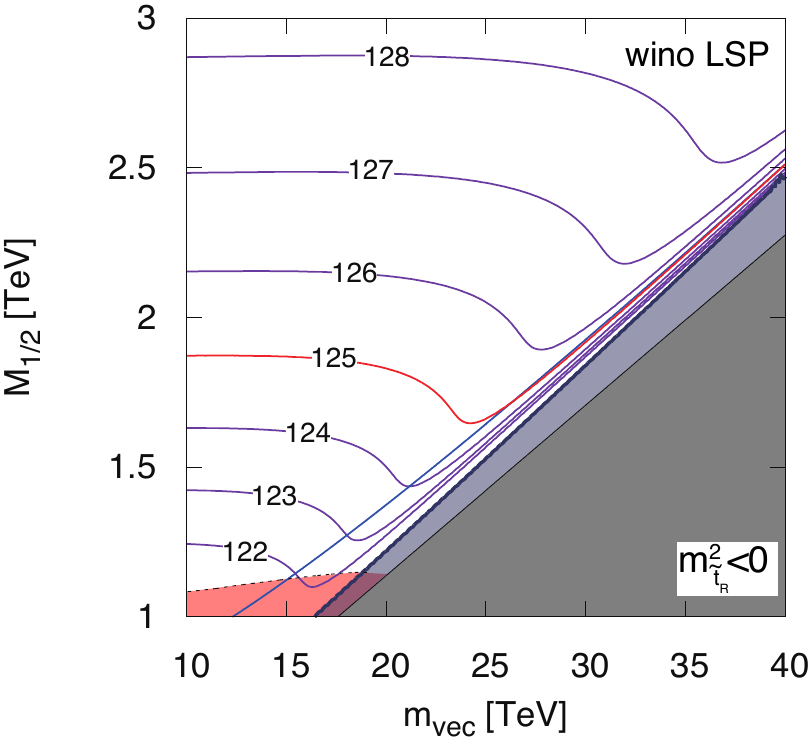}
\includegraphics[width=7cm,clip]{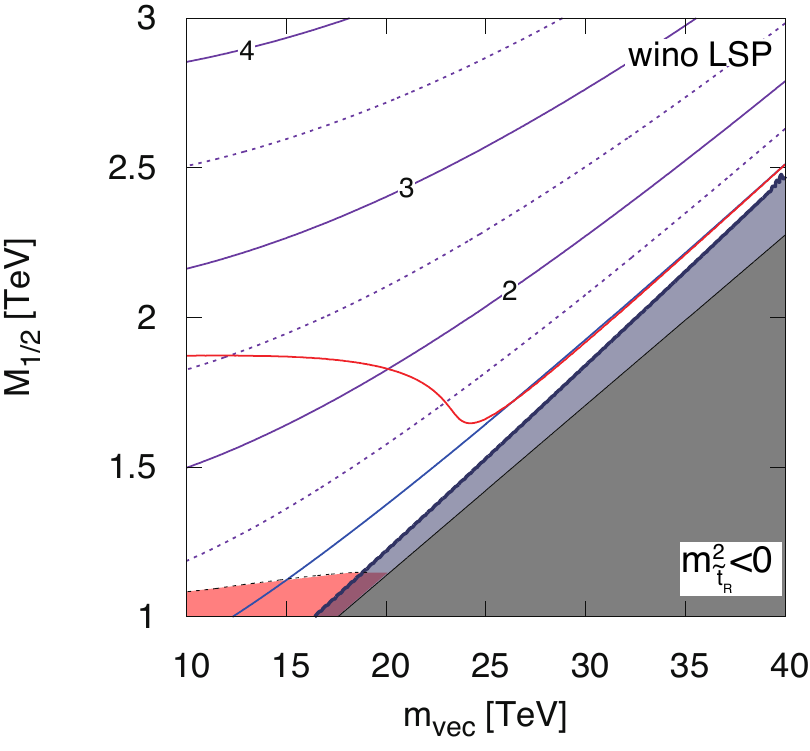}
\caption{
Dependences of light Higgs mass (left) and light stop mass (right).
An additional pair in $\mathbf{5}+\ovl{\mathbf{5}}$ is introduced.
Colored scalars in extra multiplets have large soft mass while
non-colored ones do not. Non-universal gaugino masses are also assumed.
The gray shaded region is excluded by tachyonic stops and the navy shaded region 
corresponds to the stop LSP. The gluino mass is below 1.9~TeV in the red shaded region.}
\label{fig:NUGMresult}
\end{figure}

Another possibility to avoid the stau LSP is the non-universal
gaugino mass condition at the GUT scale.\footnote{
The non-universal gaugino mass has been discussed in many situations,
for instance, in the focus point gaugino mediation scenario~\cite{Yanagida:2013ah}.
In this scenario, the observed Higgs mass is realized
without tuning by virtue of the non-universal gaugino mass.
Furthermore, the gluino and stop are accessible at the LHC
in this scenario with extra vector-like matters~\cite{Yanagida:2014cxa}.}
If the bino is heavier than
the other gauginos, we can easily shift all sfermion masses up.
The neutral wino and higgsino are the candidates for the LSP in
this setup.  We assume the initial condition for gaugino masses as
follows:
\eqs{
M_1 = 1.5 M_{1/2} \, , ~~
M_2 = 0.5 M_{1/2} \, , ~~
M_3 = M_{1/2} \, .
}
Here, the gaugino mass ratios are just assumed to raise the
right-handed stau mass. The numerical results under this boundary
condition are shown in \cref{fig:NUGMresult}.  In this figure, the LSP
dominates the neutral component of the wino in the whole unshaded
region.  From this figure, we see that the stop mass around 1~TeV, the
observed Higgs mass, and the wino LSP can be realized.

\begin{table}[h]
  \begin{center}
  \caption{Benchmark points}
  \begin{tabular}{c|cccc}
    \hline\hline
    Models & 1 & 2 & 3 & 4 \\
    \hline
    $m_{\text{vec}}$~[TeV] & 15 & 30 & 25 & 30 \\
    $M_{1/2}$~[TeV] & 1.80 & 1.96 & 1.80 & 1.90 \\
    \hline
    $m_h$ ~[GeV]& 125.2 & 125.0 & 125.9 & 124.3 \\
    $m_{\widetilde t_{1,2}}$ ~[GeV]& 2153, 2609 & 789, 1783 & 820, 1839 & 889, 1404 \\
    $m_{\widetilde b_{1,2}}$ ~[GeV]& 2585, 2763 & 1740, 1770 & 1814, 1979 & 1321, 1597 \\
    $m_{\widetilde \tau_{1,2}}$ ~[GeV]& 464, 886 & 425, 1196 & 417, 1093 & 698, 851 \\
    $m_{\widetilde u_{L,R}}$ ~[GeV]& 2884, 2790 & 2105, 1795 & 2248, 2016 & 1661, 1684 \\
    $m_{\widetilde d_{L,R}}$ ~[GeV]& 2884, 2784 & 2106, 1790 & 2249, 2010 & 1663, 1622 \\
    $m_{\widetilde e_{L,R}}$ ~[GeV]& 889, 481 & 1199, 446 & 1102, 460 & 705, 855 \\
    $m_{\widetilde g}$ ~[GeV]& 3082 & 3168 & 2954 & 3063 \\
    $m_{\widetilde\chi_1^0}$ ~[GeV]& 594 & 644 & 362 & 576 \\
    $m_{\widetilde\chi_2^0}$ ~[GeV]& 1134 & 1221 & 375 & 949 \\
    $m_{\widetilde\chi_3^0}$ ~[GeV]& 2121 & 1669 & 597 & 1619 \\
    $m_{\widetilde\chi_4^0}$ ~[GeV]& 2124 & 1676 & 1136 & 1621 \\
    $A_t$~[GeV] & -3063 & -3335 & -3063 & -3035 \\
    \hline
  \end{tabular}
  \label{tab:benchmark}
  \end{center}
  \end{table}
  \begin{table}[h]
  \begin{center}
  \caption{mSUGRA examples}
  \begin{tabular}{c|cc}
    \hline\hline
    Examples & 1 & 2 \\
    \hline
    $m_0$~[TeV] & 0.0 & 1.0 \\
    $M_{1/2}$~[TeV] & 1.9 & 1.0 \\
    $A_0$~[TeV] & 0.0 & -3.0 \\
    \hline
    $m_h$ ~[GeV] & 125.4 & 125.2 \\
    $m_{\widetilde t_{1,2}}$ ~[GeV]& 2959, 3466 & 1234, 1889 \\
    $m_{\widetilde b_{1,2}}$ ~[GeV]& 3452, 3571 & 1858, 2162 \\
    $m_{\widetilde \tau_{1,2}}$ ~[GeV]& 699, 1256 & 1022, 1183 \\
    $m_{\widetilde u_{L,R}}$ ~[GeV]& 3760, 3607 & 2277, 2207 \\
    $m_{\widetilde d_{L,R}}$ ~[GeV]& 3761, 3591 & 2278, 2200 \\
    $m_{\widetilde e_{L,R}}$ ~[GeV]& 1260, 714 & 1200, 1068 \\
    $m_{\widetilde g}$ ~[GeV]& 4123 & 2288 \\
    $m_{\widetilde\chi_1^0}$ ~[GeV]& 801 & 423 \\
    $m_{\widetilde\chi_2^0}$ ~[GeV]& 1531 & 815 \\
    $m_{\widetilde\chi_3^0}$ ~[GeV]& 2434 & 1936 \\
    $m_{\widetilde\chi_4^0}$ ~[GeV]& 2438 & 1938 \\
    $A_t$~[GeV] & -3345 & -2742 \\
    \hline
  \end{tabular}
  \label{tab:mSUGRA}
\end{center}
\end{table}

Finally, we comment on the mass spectra at sample points.  The columns for
Models 1-4 in \cref{tab:benchmark} show the mass spectra at
certain points of Figs. \ref{fig:55bar}-\ref{fig:NUGMresult}, respectively.  In particular, we set the input
values for the gaugino mass and soft masses for the extra matters in order to
realize the observed Higgs boson mass.  
In \cref{tab:benchmark}, we show the light Higgs boson mass, the third-
and first-generation sfermions, gluino, neutralinos, and $A_t$ from top to bottom.
$A_t$ is estimated at the renormalization scale $\mu = 1~\text{TeV}$ in this table.
In the model points 2, 3, and 4, the light Higgs boson with a mass of 125~GeV is
realized with 800-1000~GeV stop and roughly 3~TeV gluinos.

In order to compare our results to the MSSM with the use of \texttt{Spheno}, 
we show two examples of mass spectra with the minimal supergravity (mSUGRA) 
initial condition in \cref{tab:mSUGRA}.
We take $\tan \beta = 10$ and $\mathrm{sign}~\mu_H =+1$, again.
To see the difference from the mass spectra in the presence of the extra matters, we set an initial condition as in the gaugino mediation scenario, $m_0 = A_0 = 0$, and $M_{1/2} = 1.9~\textrm{TeV}$, in the first column.
In the absence of extra matters, the stop is much heavier than 1~TeV at low energy since there is no 
cancellation of the one-loop RGE effects from the gaugino masses.
The second column shows the mass spectrum in the context of the large A-term scenario.
Even in this case, in order to explain the 125~GeV Higgs boson and the 1~TeV stop simultaneously, 
the gluino cannot be very heavy na\"ively since the gluino mass dominates the RGEs for soft parameters.

\section{Conclusions and Discussion \label{sec:conclusion}}

In this study, we have explored the possibilities that the theories
with a heavy gluino predict a lighter stop mass ($\sim 1$~TeV) and the
observed Higgs mass ($ \sim 125~\text{GeV}$).  We have introduced
additional vector-like matter in the $SU(5)$ complete multiplets.
In particular, we have analyzed the cases with a
$\mathbf{5}+\ovl{\mathbf{5}}$ pair.  If we set the initial mass
parameters for gaugino masses and extra vector-like matters to be
above a few TeV and 20~TeV, respectively, we get the stop with a mass
of about 1~TeV and the observed Higgs mass.

The LSP in the scenario is model dependent. The stau is the LSP in the
MSSM when we impose the GUT relation of gaugino masses and
$m_{H_u}^2=m_{H_d}^2=0$ as an initial condition. Such cases may be
viable if the $R$-parity is broken or the LSP is the axino. We also found
that when $m_{H_u}^2=m_{H_d}^2>0$ is assumed at the GUT scale, the higgsino-like
neutralino is the LSP near the boundary for no EWSB.  In the
non-universal gaugino mass scenario in which the bino is heavier than the
other gauginos, the neutral wino is the LSP.

Hierarchical structure in soft parameters is motivated by the gaugino
mediation mechanism.  The soft parameters for gauginos and scalars in
vector-like extra matters are assumed to be non-zero values since they
are coupled with the SUSY-breaking brane directly.  The scalars localized on
our brane (squarks and sleptons) obtain no soft masses at tree level.
The hierarchical structure gives a negative contribution to the
scalar soft mass squared for squarks and sleptons.  The large gaugino
mass at the input scale leads to large values for the A-terms.  As a
result, we found that the observed Higgs mass could be explained in
scenarios with a heavy gluino and light stop.

We note that we have determined the initial conditions for
the soft parameters by hand.
The essential ingredient of this work is the assumption that there is a large hierarchy between the gaugino mass and soft masses for extra matters.
This assumption seems to be reasonable since the gaugino masses can be suppressed by some sort of chiral symmetry.
We have also assumed non-universal gaugino masses or non-universal soft mass for components of a $\mathbf{5}+\ovl{\mathbf{5}}$ pair.
This is needed to construct concrete mediation
models giving specific boundary conditions for the soft parameters.  The
model building, however, is beyond the scope of this study, and thus
we leave it for future work.

Finally, the introduction of the extra matter leads to fruitful
phenomenology. While the FCNC processes are suppressed in our setup,
non-vanishing electric dipole moments (EDMs) for the electron and nucleons
may be predicted at one-loop level if the $b$-term in the Higgs
potential is non-zero at the GUT scale. Even if the $b$-term at the GUT
scale is zero, the EDMs may get the two-loop contributions by
integrating out the vector-like extra matter
\cite{Hisano:2015rna}. The gauge coupling constants at the GUT scale
are larger due to the introduction of the extra matters, and this implies
that the $X$-boson proton decay rate is enhanced
\cite{Hisano:2012wq}. Introduction of the extra matter leads to new
phenomenology which should be pursued further.

\section*{Acknowledgments}
We are grateful to J.D.~Wells for useful discussion and also comments
on our manuscript.  This work is supported by Grant-in-Aid for
Scientific Research from the Ministry of Education, Science, Sports,
and Culture (MEXT), Japan, No.~16H06492 (for J.H.). The work of
J.H. is also supported by World Premier International Research
Center Initiative (WPI Initiative), MEXT, Japan. The work of T.K. is
supported by Research Fellowships of the Japan Society for the
Promotion of Science (JSPS) for Young Scientists (No.~16J04611).

\newpage
\section*{Appendix}
\appendix

\section{Renormalization Group Equations\label{app:RGE}}
Here, we give the modification of the RGEs for soft parameters.
We follow the notation of Ref. \cite{Martin:1993zk}.
In our model, we should modify some RGEs between the input scale (GUT scale) and 1~TeV, where we integrate out the SUSY partners and vector-like multiplets.

First, we consider the modification of the RGEs for squared masses of squarks and sleptons.
This modification is divided into two parts; one is proportional to the soft masses of extra scalar fields, the other relates to the gaugino masses.
We include the scalar mass contributions from extra vector-like multiplets.
The corresponding part of the RGEs is the following:
\eqs{
\left. \diff{1}{m_r^2}{\ln\mu} \right|_{\text{Scalar-gauge}} = 2 \left( \frac{\alpha_Y}{4\pi} \right) Y_r \curlS
+ 4 \sum_A  \left( \frac{\alpha_A}{4\pi} \right)^2 C_A(r) \si_A
+ \frac{1}{4\pi^2} \left( \frac{\alpha_Y}{4\pi} \right) Y_r  \curlS' \, ,
}
where
\eqs{
\curlS & = \sum_s Y_s m_s^2 \, , \\
\curlS' & = 2 \sum_s \sum_A g_A^2 Y_s C_A(s) m_s^2 \, , \\
\si_A & = 2 \sum_s S_A(s) m_s^2 \, .
}
Here, indices ($A = 1\text{-}3$) represent the SM gauge group and we adopt the GUT normalization for $U(1)_Y$ coupling, that is $g_Y^2 = \frac35 g_1^2$.
$C_A(s)$ and $S_A(s)$ are defined in the text ( \cref{eq:RGEsforsfermion}).
The summation of $s$ runs over the degrees of freedom for the field $s$.

If we include the additional contribution from vector-like multiplets, the quantities introduced above are modified as follows:
\eqs{
\curlS & \to \curlS + \Del \curlS \, , \\
\curlS' & \to \curlS' + \Del \curlS' \, , \\
\si_A & \to \si_A + \Del \si_A, ~~ (A = 1\text{-}3) \, . \\
}
The modification parts for each quantity are given by
\eqs{
\Del \curlS & = n_{\mathbf{5}} (m_{\ovl L'}^2 - m_{D'}^2 - m_{L'}^2 + m_{\ovl D'}^2 )
+ n_{\mathbf{10}} (m_{Q'}^2 - 2 m_{\ovl U'}^2 + m_{\ovl E'}^2 - m_{\ovl Q'}^2 + 2 m_{U'}^2 - m_{E'}^2 ) \, , \\
\Del \curlS' & = - n_{\mathbf{5}} \left[ \frac32 g_2^2 + \frac{3}{10} g_1^2 \right] (m_{L'}^2 - m_{\ovl L'}^2)
+ n_{\mathbf{5}} \left[ \frac83 g_3^2 + \frac{2}{15} g_1^2 \right] ( m_{\ovl D'}^2 - m_{D'}^2 ) \\
& + n_{\mathbf{10}}\left[ \frac83 g_3^2 + \frac32 g_2^2 + \frac{1}{30} g_1^2 \right]  (m_{Q'}^2- m_{\ovl Q'}^2)
 - 2 n_{\mathbf{10}}\left[ \frac83 g_3^2 + \frac{8}{15} g_1^2 \right] (m_{\ovl U'}^2 - m_{U'}^2) \\
& + n_{\mathbf{10}} \frac65 g_1^2 (m_{\ovl E'}^2 - m_{E'}^2 ) \, , \\
}
and
\eqs{
\Del \si_1 & = \frac{n_{\mathbf{5}}}{5} g_1^2 \left[ 3 (m_{L'}^2 + m_{\ovl L'}^2) + 2 (m_{\ovl D'}^2 + m_{D'}^2) \right] \\
& + \frac{n_{\mathbf{10}}}{5} g_1^2 \left[ (m_{Q'}^2 + m_{\ovl Q'}^2) + 8 (m_{\ovl U'}^2 + m_{U'}^2) + 6 (m_{\ovl E'}^2 + m_{E'}^2) \right] \, , \\
\Del \si_2 & = n_{\mathbf{5}} g_2^2 (m_{L'}^2 + m_{\ovl L'}^2)
+ n_{\mathbf{10}} g_2^2 3 (m_{Q'}^2 + m_{\ovl Q'}^2) \, , \\
\Del \si_3 & = n_{\mathbf{5}} g_3^2 (m_{\ovl D'}^2 + m_{D'}^2)
+ n_{\mathbf{10}} g_3^2 \left[ 2 (m_{Q'}^2 + m_{\ovl Q'}^2) + (m_{\ovl U'}^2 + m_{U'}^2) \right] \, . \\
}
Here, $n_{\mathbf{5}}$ and $n_{\mathbf{10}}$, respectively, represent the numbers of the $\mathbf{5}+\ovl{\mathbf{5}}$ and $\mathbf{10}+\ovl{\mathbf{10}}$ pairs.
$Q', L', \ovl U', \ovl D'$, and $\ovl E'$ represent the additional superfields that have the same quantum numbers as the MSSM ones; on the other hand, $\ovl Q', \ovl L', U', D'$, and $E'$ represent the fields with opposite quantum number.

Including the contributions proportional to the gaugino masses, we obtain the total modification of the soft scalar mass RGEs,
\eqs{
\diff{1}{m_r^2}{\ln\mu} = \left. \diff{1}{m_r^2}{\ln\mu} \right|_{\text{MSSM}}
& + 2 \left( \frac{\alpha_Y}{4\pi} \right) Y_r \Del \curlS
+ 4 \sum_A  \left( \frac{\alpha_A}{4\pi} \right)^2 C_A(r) \Del \si_A
+ \frac{1}{4\pi^2} \left( \frac{\alpha_Y}{4\pi} \right) Y_r  \Del \curlS' \\
& + \sum_A 24 C_A(r) \left(\frac{\alpha_A}{4\pi}\right)^2 |M_A|^2 (n_5 + 3 n_{10}) \, .
}
Here, we ignore the contributions from Yukawa couplings between extra vector-like pairs and MSSM Higgs doublets.

The two-loop beta functions in the presence of $SU(5)$ complete vector-like multiplets are given in Ref.~\cite{Ghilencea:1997mu}.
The two-loop RGEs for MSSM Yukawa couplings are corrected via the anomalous dimension for the MSSM Higgs doublets,
\eqs{
\diff{1}{Y_u}{\ln\mu} & = \left. \diff{1}{Y_u}{\ln\mu} \right|_{\text{MSSM}}
+ \frac{Y_u}{(16\pi^2)^2} (n_5 + 3 n_{10}) \left( \frac{16}{3} g_3^4 + 3 g_2^4 + \frac{13}{15} g_1^4 \right) \, , \\
\diff{1}{Y_d}{\ln\mu} & = \left. \diff{1}{Y_d}{\ln\mu} \right|_{\text{MSSM}}
+ \frac{Y_d}{(16\pi^2)^2} (n_5 + 3 n_{10}) \left( \frac{16}{3} g_3^4 + 3 g_2^4 + \frac{7}{15} g_1^4 \right) \, , \\
\diff{1}{Y_e}{\ln\mu} & = \left. \diff{1}{Y_e}{\ln\mu} \right|_{\text{MSSM}}
+ \frac{Y_d}{(16\pi^2)^2} (n_5 + 3 n_{10}) \left( 3 g_2^4 + \frac{9}{5} g_1^4 \right) \, . \\
}
\newpage
\bibliography{ref}
\end{document}